\documentclass[a4paper]{article}

\usepackage[english]{babel}
\usepackage[utf8]{inputenc}
\usepackage{booktabs}
\usepackage{amsmath}
\usepackage{graphicx}
\usepackage[colorinlistoftodos]{todonotes}
\usepackage[numbers,round]{natbib}
 \usepackage{setspace}
\title{Tweedie distributions for fitting semicontinuous health care utilization cost data}

\author{Christoph Kurz}

\date{\today}

\begin{document}
\maketitle

\begin{abstract}
\noindent We explore a statistical distribution that can simultaneously model the probability of zero outcome for non-users of health care utilization and continuous costs for users. We compare this distribution to other commonly used models on example data and show that it fits cost data well and has some appealing properties that provide flexible use.
\end{abstract}
\onehalfspacing
\section{Introduction}
In modeling cost data of health care utilization, the non-negative response variable
is usually right-skewed with a positive probability of zero outcome for non-users. Such variables are called \emph{semicontinuous}~\cite{min2002modeling} and pose a number of problems: because of the point mass at zero, common models involving the Gamma or lognormal distributions have difficulty with such a mixture of discrete and continuous values. A popular way to account for this in the generalized linear models (GLM) framework are two-part models~\cite{duan1983comparison}, which combine a binary model for the dichotomous event of having either zero or positive values with a continuous model for those having positive values. This complements a two-stage decision process which can be inadequate because the two decisions are usually not made independently (Winkelmann~\cite{winkelmann2004health} and Van Ophem~\cite{van2011frequency} discuss this for the case of physicians visits). % and can therefore be translated to costs as well).
Another more simple model, using a single distribution, is the Tobit model~\cite{tobin1958estimation}. This model is based on a zero-truncated Normal distribution but cannot handle \emph{excess zeros}, i.e. the presence of more zeros in the data than what would be expected by the underlying distribution. % better: no natural support for point mass at zero?
In this linear regression setting, constant variance is assumed, which is also inadequate for cost data.
%As analysts favor simple models that are easy to interpret, we restrict ourselves to these alternatives. For an overview of other, more specialized methods, we refer to~\cite{mihaylova2011review}.

In this paper, we consider a single distribution GLM for cost data that can simultaneously model the zeros and continuous positives outcomes. The number of excess zeros can be arbitrarily high while still providing good support for the positive costs. Variance can be specified as some power of the mean. This model, based on the family of Tweedie densities~\cite{jorgensen1997theory} has already shown to perform well in the case of rainfall precipitation~\cite{dunn2004occurrence} and insurance premium~\cite{smyth2002fitting}. To our knowledge, the Tweedie densities have not been used in health economic cost data modeling before. In the following we compare the Tweedie model to the Two-part (Binomial/Gamma) and Tobit regarding parameter estimates, model fit and prediction error. As analysts favor simple models that are easy to interpret, we restrict ourselves to these alternatives. For an overview of other, more specialized methods, we refer to~\cite{mihaylova2011review}.

The rest of this paper is structured as follows: Section 2 describes the properties of the Tweedie family of distributions and explains the proposed model; Section 3 outlines the data;
Section 4 compares the Tweedie with the Two-part and Tobit model on these data; and Section 5 discusses the results.

Code and data to reproduce all analyses are available on the author's github page~\cite{github}.

\section{Tweedie Family Densities}
We outline the model used in this paper as a special case of exponential dispersion models (EDMs)~\cite{jorgensen1997theory}. This class of models is a broad family of distributions defined by the form
$$
f(y|\theta, \phi) = a(y, \phi) \exp \left[ \frac{y\theta - \kappa(\theta)}{\phi} \right],
$$
where %$\mu$ is the mean of the distribution and 
both the normalizing functions $a(\cdot)$ and $\kappa(\cdot)$ are known. $\theta$ is the natural parameter and $\phi > 0$ is called the dispersion parameter. Mean $\mu$ and variance of a random variable $Y$ from an EDM are given by $E(Y)=\mu=\kappa'(\mu)$ and $\text{Var}(Y)=\kappa''(\mu)\phi$, respectively.
The Tweedie family of distributions corresponds to special cases of EDMs where the power mean-variance relationship is characterized by $\text{Var}(\mu)=\phi\mu^p$ for $p\not\in(0,1)$.
The Tweedie family includes a number of familiar distributions, e.g. Normal ($p=0$), Poisson ($p=1$), Gamma ($p=2$) and inverse Gaussian ($p=3$).

For the case of cost data modeling the choice $p\in(1,2)$ is the most interesting one and our main focus here, because of its support for semicontinuous outcomes. Tweedie distributions in this range of $p$ belong to the so-called compound Poisson-Gamma distributions~\cite{jorgensen1997theory}. Let $M\sim \text{Pois}(\lambda)$ be a Poisson random variable, and let $X_i\stackrel{iid}{\sim} \text{Gamma}(\alpha, \beta)$ be Gamma distributed with $M \perp X_i$, then a random variable $Z$, defined by
$$
Z=\begin{cases}
0,  & \text{if }M=0,\\
X_1+X_2+...+X_M, & \text{if }M=1,2,...
\end{cases},
$$ 
follows a compound Poisson-Gamma distribution, i.e. is a Poisson sum of Gamma random variables. If $M=0$ then $Z=0$, thus allowing for a probability mass at zero for non-users, where $\text{Pr}(Z=0)=\exp(-\lambda)$. If $M>0$, then $Z$ is the sum of $M$ iid Gamma random variables, so conditional on $M$, $Z|M\sim \text{Gamma}(M\alpha, \beta)$, resulting in a continuous distribution for the positive outcome. With $M=m$ the distribution for $z>0$ is therefore given as:
$$
f(z|\lambda, \alpha, \beta) = \frac{\lambda^m \exp(-\lambda)}{m!} \frac{z^{m\alpha-1}\exp(-z/\beta)}{\beta^{m\alpha}\Gamma (m\alpha)}.
$$
These parameters $\lambda, \alpha$ and $\beta$ are related to the Tweedie distribution parameters $\mu, \phi$, and $p$ by:
$$
\lambda = \frac{\mu{2-p}}{\phi(2-p)}, \quad
\alpha = \frac{2-p}{\alpha -1}, \quad
\beta = \phi(p-1)\mu^{p-1}.
$$
Recovering the underlying marginal distribution of $Z$ results in a non-closed form expression for the normalizing function $a(\cdot)$, based on Wright's generalized Bessel function $W(\cdot, \cdot, \cdot)$~\cite{dunn2004occurrence, zhang2013likelihood}:
$$
a(z,\phi)=\frac{1}{y}W(z,\phi,p).
$$
Dunn and Smythe~\cite{dunn2008evaluation} show that this function is strictly convex and can be approximated by Stirling's formula for the Gamma function and a Fourier inversion method for the infinite series.

Because Tweedie distributions also belong to the exponential family of distributions, they can be used in the GLM framework~\cite{mccullagh1989generalized}. Besides the ability to model exact zeros and continuous outcomes, the idea that positive total costs are sums of smaller costs provide an intuitive appeal: %(as outlines by Duan~\cite{dunn2004occurrence} for precipitation). 
$Z$ is the total amount of expenses in a given period, $M$ the number of utilization events, and $X_i$ the expenses of the $i$-th event.
In the following we show that the Tweedie distributions fit health care utilization cost data very well.

\section{Data} 
We use data from the RAND Health Insurance Experiment (RAND HIE).
This US study %study took place from 1974 to 1982 in the United States and 
measured health care costs, among other outcomes, of people randomly assigned to different kinds of plans.
Because of the random assignment, the reliability of health insurance coverage, and the availability of important variables for this application, these data provide an accurate base for cost modeling in our case. 

As outcome variable we use the total costs, consisting of outpatient, drug, supply, psychotherapy and inpatient expenses. We use covariates commonly considered to determine health care utilization. Among the socio-economic characteristics are age, gender, race, the logarithm of family income, LINC, the number of physical limitations, PHYSLM, the number of chronic diseases, DISEA, the logarithm of family size, LFAM, the education of household head in years, EDUCDEC, and a dummy variable indicating self-rated health as good, HLTHG. Insurance specific variables included are the log coinsurance rate plus 1, LOGC, a dummy for the individual deductible plan, IDP, the log of the participation incentive payment, LPI, and a maximum expenditure function, FMDE. A more detailed description of the data set and the variables can be found in~\cite{deb2002structure}.

We only use the first year of observation for each individual 18 years and older ($n=3301$). There are 18.1\% zero observations for the costs with a mean of 206.80 and a maximum of 17730.  

\section{Application}
In this section, we apply the Two-part, the Tweedie and the Tobit model on the RAND HIE data.
The Two-part model involves two estimations: first, we decide whether someone has zero or non-zero costs using a logistic regression. Second, conditional on having non-zero costs, we apply a Gamma model with log-link to the positive outcome. We also use the log-link for the Tweedie. The Tobit model features an underlying Normal distribution truncated at zero.
For a more detailed description and justification of the Two-part and Tobit models see~\cite{min2002modeling} and~\cite{mihaylova2011review}.
The aim of this application is to show how model choice affects estimates, model fit and prediction in the case of semicontinuous health care cost data. No causal effects will be revealed in this study.

Table~1
%\ref{tab:1} 
presents the parameter estimation results for the discussed models.
While the parameter estimates of the Tobit model are on a completely different scale, the Tweedie and the Gamma part of the Two-part model are very similar: all estimates share the same sign and have comparable values, leading to the same conclusions. 
Looking at the standard errors, we see that both Tweedie and Gamma estimations lead to very similar estimated standard errors, with the Gamma slightly underestimating the standard error.
Furthermore, the log-likelihood of the Tweedie and Two-part are almost identical, suggesting comparable model fit: The Tweedie log-likelihood is -18874.4, while the Two-part has a combined log-likelihood of -18597.9 . The log-likehood of the Tobit model is slightly lower with a value of -21809.2. 
If we plot the true and estimated quantiles of the cost outcome for all distributions against each other, the Two-part has better model fit for the lower quantiles, while the Tweedie has slightly higher support for upper quantiles. See Figure~1 for these Q-Q plots. This is likely because of the heavier tails of the Tweedie distribution. The Tobit model fits the central quantiles badly.

For the last comparison we randomly split the data into training set ($n=2801$) and test set ($n=500$). We build each model only on the training set and then evaluate on the test set. As comparison metric we chose root mean squared error (RMSE), defined by
$$
\operatorname{RMSE}=\sqrt{\frac{1}{n}\sum_{i=1}^n (\hat y_i - y_i)^2},
$$ 
where $\hat y$ denotes the estimate and $y$ the true value. Figure~2 plots the estimated total health care utilization costs against the true ones in the test set. Again, Two-part and Tweedie produce visually comparable estimates. The Tweedie model has the lowest RMSE with a value of 467.67, followed by the Two-part with 467.71 and the Tobit with 471.58.

The estimated value for the mean-variance power parameter in the Tweedie model is $p=1.719$. Figure~3 shows the mean-variance plots for all 5\% quantiles for the Tweedie model on the example data.

\section{Discussion}
This paper explores a single distribution GLM based on the Tweedie family of distributions for semicontinuous cost data. This model is comparable in model fit to the two-part Binomial/Gamma model but only includes a one-stage decision process, making it easier to interpret. The Tweedie model outperforms the Tobit model as the popular single distribution model for non-negative continuous data with a support for exact zeros. Thus, it provides an interesting alternative for modeling of health care utilization cost data as it has a natural support for cases where no utilization has occurred and for those it has. The theoretical justification is given since for the discussed case where the power parameter $p\in(1,2)$, the Tweedie model can be explained as a Poisson sum of Gamma distributions. There, the number of utilization events is expressed by a Poisson distribution, and the amount of each utilization by a Gamma distribution. 

While the theory of the Tweedie families is known for more than 20 years, only recently software packages that allow easy handling of these distributions became available~\cite{zhang2013likelihood, cplm}. 
Further research should explore the usefulness of Tweedie distributions with $p>2$ as they provide similar shape to the Gamma but support heavier tails. Tweedie models in this range may be an attractive alternative for the continuous part of a Two-part model or for cases without exact zeros and support a more flexible mean-variance relationship. The estimated mean-variance power parameter $p=1.719$ may not appropriately reflect the true relationship. The fixed $p=2$ in Gamma models is still too low, but values of $p$ in the range of 2.2 to 2.3 seem more realistic.

Swallow et al~\cite{swallow2015bayesian} show in an ecological setting that a Bayesian hierarchical model based on the Tweedie densities provide further flexibility and removes this need to make strong assumptions about mean-variance relationships a priori. Such an hierarchical extension may further be useful to account for correlated effects by repeated measurement of individuals, for example in clinical trial settings or claims data.

\section*{Acknowledgements}
The author thanks Rolf Holle and Irena Cenzer for helpful comments on the manuscript.

%todo: 
%does not predict exact zeros
\begin{table}[ht]\label{tab:1}
	\centering
	\begin{tabular}{lrrrrrrrr}
		\toprule
		%& Tobit & Tweedie  & 
		& \multicolumn{2}{c}{Tobit}
		& \multicolumn{2}{c}{Tweedie}		
		& \multicolumn{4}{c}{Two-part}		
		 \\
		\cmidrule(rl){2-3}\cmidrule(rl){4-5}
		\cmidrule(rl){6-9}
		 
        & & & & & \multicolumn{2}{c}{Binomial} 
        & \multicolumn{2}{c}{Gamma} 
        \\
        \cmidrule(rl){6-7}
        \cmidrule(rl){8-9}
		& Est. & SE & Est. & SE  & 
		Est. & SE & Est. & SE  
		\\\midrule		
%		\hline
%		& tob & tob.se & tweedie & tweedie.se & h1 & h1.se & h2 & h2.se \\ 
		%\hline
%		Intercept & -428.481 & 237.275 & 4.966 & 0.552 & -0.731 & 0.408 & 5.691 & 0.530 \\ 
%		age & 4.369 & 2.259 & 0.007 & 0.005 & 0.015 & 0.005 & 0.004 & 0.005 \\ 
%		disea & 11.985 & 3.408 & 0.014 & 0.008 & 0.044 & 0.008 & 0.007 & 0.007 \\ 
%		physlm & 422.796 & 69.200 & 0.687 & 0.153 & 0.319 & 0.156 & 0.664 & 0.148 \\ 
%		logc & -48.446 & 34.881 & -0.050 & 0.080 & -0.185 & 0.065 & -0.020 & 0.076 \\ 
%		idp & -17.088 & 67.918 & 0.046 & 0.156 & -0.063 & 0.130 & 0.052 & 0.150 \\ 
%		lpi & -3.174 & 11.147 & -0.027 & 0.025 & 0.004 & 0.022 & -0.027 & 0.024 \\ 
%		fmde & 3.705 & 20.798 & 0.020 & 0.048 & 0.007 & 0.039 & 0.020 & 0.045 \\ 
%		linc & 37.578 & 23.806 & 0.066 & 0.056 & 0.098 & 0.038 & 0.031 & 0.055 \\ 
%		lfam & -26.072 & 46.206 & -0.010 & 0.107 & 0.051 & 0.088 & -0.002 & 0.101 \\ 
%		female & 270.586 & 52.504 & 0.392 & 0.121 & 0.904 & 0.104 & 0.265 & 0.113 \\ 
%		black & -336.386 & 78.705 & -0.349 & 0.182 & -1.044 & 0.130 & -0.110 & 0.178 \\ 
%		educdec & -0.162 & 9.156 & -0.014 & 0.021 & 0.045 & 0.018 & -0.020 & 0.020 \\ 
%		hlthg & -23.676 & 54.025 & -0.057 & 0.125 & 0.106 & 0.105 & -0.074 & 0.116 \\ 
%		hlthf & 74.705 & 93.280 & 0.079 & 0.210 & 0.046 & 0.185 & 0.065 & 0.204 \\ 
Intercept & -212.276 & 118.391 & 4.253 & 0.543 & -0.732 & 0.405 & 4.972 & 0.522 \\ 
age & 2.274 & 1.135 & 0.007 & 0.005 & 0.015 & 0.005 & 0.004 & 0.005 \\ 
disea & 6.319 & 1.695 & 0.015 & 0.008 & 0.044 & 0.008 & 0.008 & 0.007 \\ 
physlm & 214.712 & 34.484 & 0.688 & 0.150 & 0.321 & 0.156 & 0.664 & 0.145 \\ 
logc & -23.994 & 17.530 & -0.046 & 0.079 & -0.184 & 0.065 & -0.016 & 0.076 \\ 
idp & -7.057 & 34.130 & 0.051 & 0.155 & -0.061 & 0.130 & 0.057 & 0.149 \\ 
lpi & -1.806 & 5.612 & -0.028 & 0.025 & 0.005 & 0.022 & -0.028 & 0.024 \\ 
fmde & 1.728 & 10.453 & 0.018 & 0.047 & 0.007 & 0.039 & 0.018 & 0.045 \\ 
linc & 18.113 & 11.951 & 0.066 & 0.055 & 0.096 & 0.038 & 0.032 & 0.055 \\ 
lfam & -12.640 & 23.245 & -0.013 & 0.106 & 0.054 & 0.088 & -0.008 & 0.101 \\ 
female & 138.240 & 26.417 & 0.397 & 0.120 & 0.904 & 0.104 & 0.270 & 0.112 \\ 
black & -166.908 & 39.181 & -0.348 & 0.179 & -1.040 & 0.128 & -0.112 & 0.174 \\ 
educdec & 0.150 & 4.556 & -0.011 & 0.021 & 0.044 & 0.018 & -0.017 & 0.019 \\ 
hlthg & -17.324 & 25.583 & -0.067 & 0.116 & 0.102 & 0.100 & -0.082 & 0.108 \\ 

		\hline
				p 		& \multicolumn{2}{c}{}
				& \multicolumn{2}{c}{1.719}		
				& \multicolumn{2}{c}{}
				& \multicolumn{2}{c}{} \\
				\emph{dispersion} 		& \multicolumn{2}{c}{1}
				& \multicolumn{2}{c}{9.518}		
				& \multicolumn{2}{c}{1}
				& \multicolumn{2}{c}{7.629} \\
		\hline
		\emph{log-likelihood} 		& \multicolumn{2}{c}{-21809.21}
		& \multicolumn{2}{c}{-18874.42}		
		& \multicolumn{2}{c}{-1371.20}
		& \multicolumn{2}{c}{-17225.84}
	\end{tabular}
	\caption{comparison of estimates}
\end{table}

\begin{figure}
\centering
\includegraphics[width=0.99\textwidth]{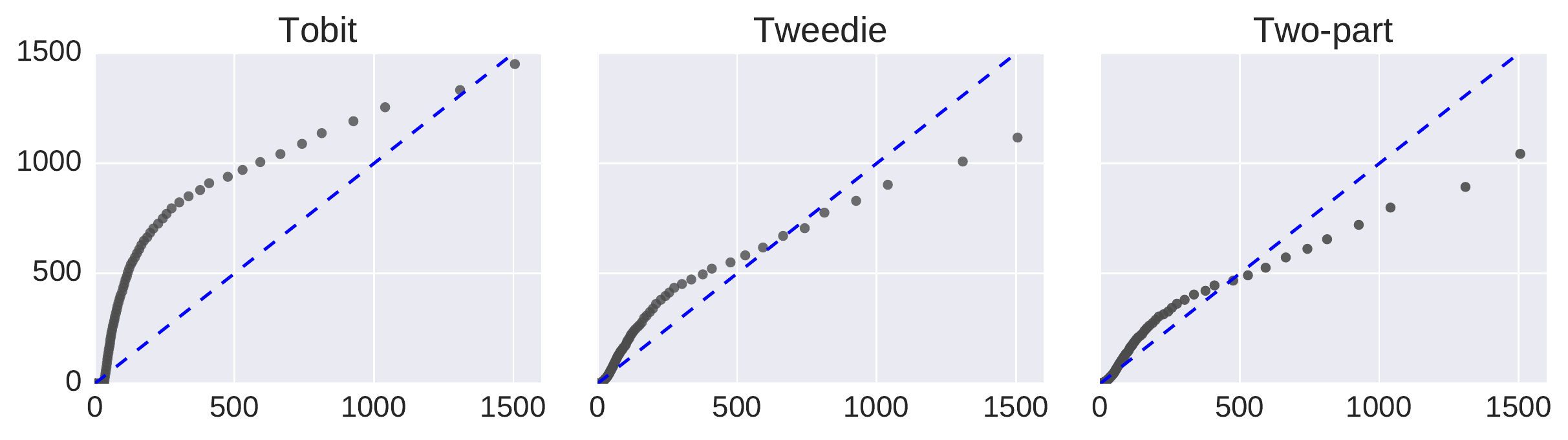}
\caption{\label{fig:qqplot}Q-Q plots for true and estimated quantiles of total health care utilization in the RAND HIE data for all models.
	Because of heavy outliers we do not show the last percentile. Quantile values closer to the dashed line represent a better match of empirical and estimated distributions.}
\end{figure}

\begin{figure}
	\centering
	\includegraphics[width=0.99\textwidth]{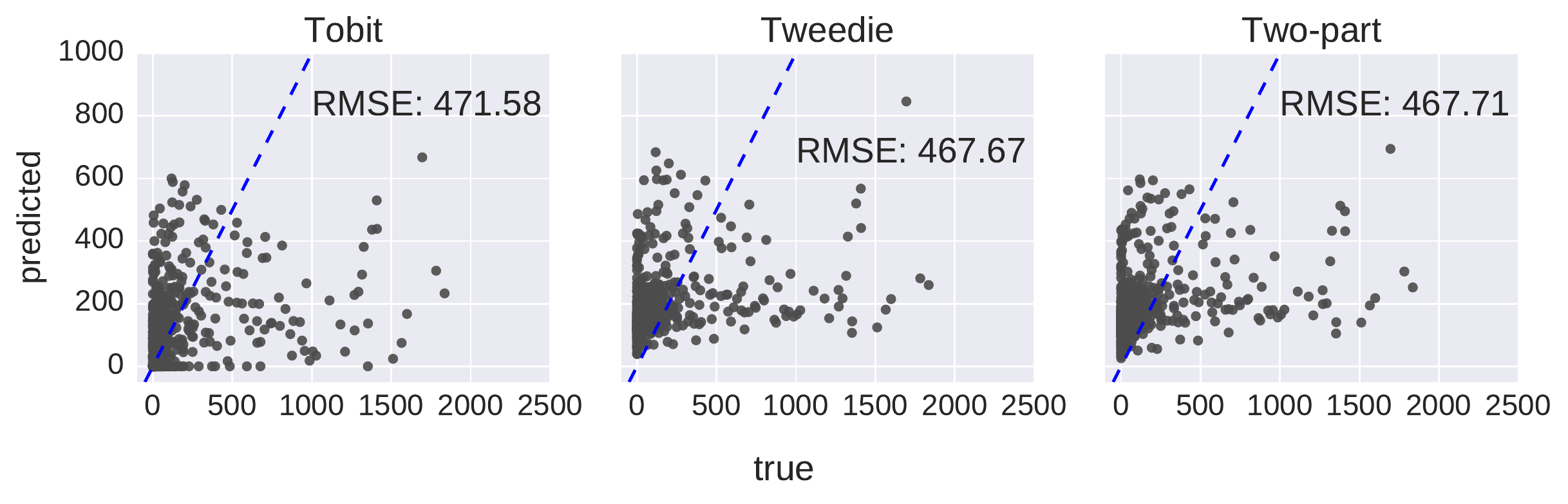}
	\caption{\label{fig:rmseplot}Plots of predicted versus true costs for total health care utilization for a test subset of the RAND HIE data.
		Predictions closer to the dashed line are more accurate. Some outliers exceeding the axis limits are not shown.}
\end{figure}

\begin{figure}
	\centering
	\includegraphics[width=0.7\textwidth]{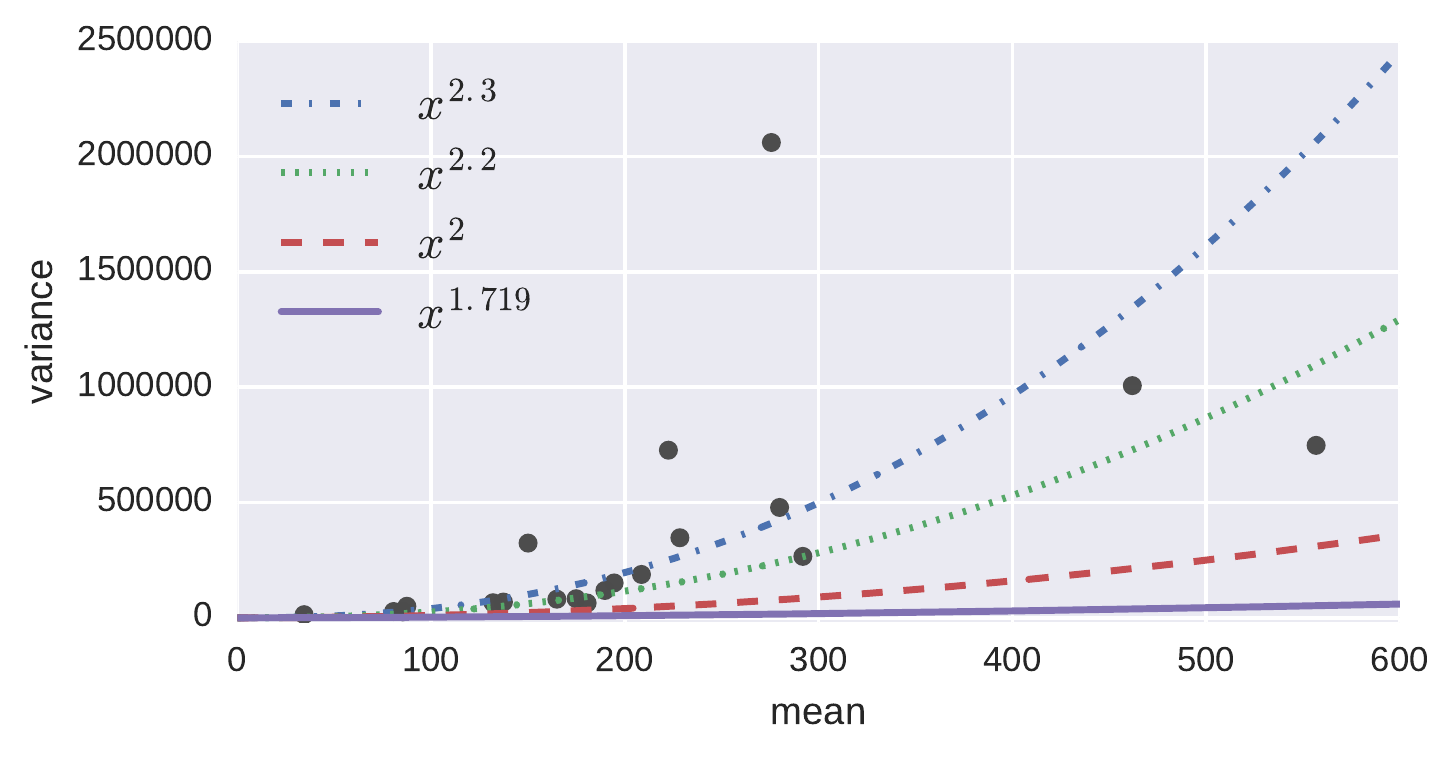}
	\caption{\label{fig:rmseplot}Mean-variance plots for all 5\% quantiles for the Tweedie model. The solid line represents the estimated value for the mean-variance power parameter $p=1.719$. Other values are plotted for comparison.}
\end{figure}

\newpage

\bibliographystyle{abbrv}
\bibliography{paper_tweedie}

\end{document}